# Van der Waals epitaxy of Weyl-semimetal $T_d$-WTe$_2$

*Alexandre Llopez[1], Frédéric Leroy[1], Calvin Tagne-Kaegom[2], Boris Croes[1,3], Adrien Michon[4], Chiara Mastropasqua[4], Mohamed Al Khalfioui[4], Stefano Curiotto[1], Pierre Müller[1], Andrés Saùl[1], Bertrand Kierren[2], Geoffroy Kremer[2], Patrick Le Fèvre[5,6], François Bertran[5], Yannick Fagot-Revurat[2], Fabien Cheynis[1]\**

[1] Aix Marseille Univ, CNRS, CINAM, AMUtech, Marseille, France

[2] Institut Jean Lamour, UMR 7198, CNRS-Université de Lorraine, Campus ARTEM, 2 allée André Guinier, BP 50840, 54011 Nancy, France

[3] Université de Strasbourg, CNRS, IPCMS, UMR 7504, 23 Rue du Loess Bâtiment 69, 67000 Strasbourg, France

[4] CRHEA, Université Côte d'Azur, CNRS, Rue Bernard Grégory, 06560 Valbonne, France

[5] Université Paris-Saclay, Synchrotron SOLEIL, L'Orme des Merisiers, Départementale 128, 91190 Saint-Aubin, France

[6] Univ Rennes, CNRS, IPR - UMR 6251, F-35000 Rennes, France

KEYWORDS: van der Waals epitaxy, $T_d$-WTe$_2$, monolayer, bilayer, transition metal dichalcogenide, nucleation and growth, angle-resolved photoemission spectroscopy, DFT calculations, electronic structure.

ABSTRACT:

Epitaxial growth of WTe$_2$ offers significant advantages, including the production of high-quality films, possible long range in-plane ordering and precise control over layer thicknesses. However, the mean island size of WTe$_2$ grown by molecular beam epitaxy (MBE) in litterature is only a few tens of nanometers, which is not suitable for an implementation of devices at large lateral scales. Here we report the growth of Td -WTe$_2$ ultrathin films by MBE on monolayer (ML) graphene reaching a mean flake size of ≃110nm, which is, on overage, more than three time larger than previous results. WTe$_2$ films thicker than 5nm have been successfully synthesized and exhibit the expected Td-phase atomic structure. We rationalize epitaxial growth of Td-WTe$_2$ and propose a simple model to estimate the mean flake size as a function of growth parameters that can be applied to other transition metal dichalcogenides (TMDCs). Based on nucleation theory and Kolmogorov-Johnson-Meh-Avrami (KJMA) equation, our analytical model supports experimental data showing a critical coverage of 0.13ML above which WTe$_2$ nucleation becomes negligible. The quality of monolayer WTe$_2$ films is demonstrated from electronic band structure analysis using angle-resolved photoemission spectroscopy (ARPES) in agreement with first-principle calculations performed on free-standing WTe$_2$ and previous reports. We evidence electron-pockets at the Fermi level indicating a *n*-type doping of WTe$_2$ with an electron density of $n = 2.0 \pm 0.5 \times 10^{12}\ cm^{-2}$ for each electron pocket.

INTRODUCTION

Transition metal dichalcogenides (TMDCs) are two-dimensional (2D) layered materials that have recently garnered a huge scientific interest due to their unique thickness-dependent properties and their potential for various applications in electronics, photonics, and energy[1,2]. Among TMDCs, WTe$_2$ is the only material with the T$_d$ phase (space group: *Pmn2$_1$*) as most energetically favorable[3] polytype. T$_d$-phase is characterized by an inversion symmetry breaking[4,5] leading to a ferroelectric behavior down to the bilayer[6]. It has been anticipated to be a bulk type-II Weyl semimetal[7] and observations of Fermi arcs strongly supports this prediction[8,9]. In the monolayer regime[4,5], the usually-labelled 1T'-WTe$_2$ has been predicted and experimentally confirmed to be a quantum spin-Hall insulator (QSHI) characterized by edge conducting states and band gap opening due to spin-orbit coupling (SOC)[10,11]. In most reported studies, mono or multilayers TMDCs are obtained through mechanical or chemical exfoliation from TMDC bulk crystals, which is not transferable to technological applications. Thin TMDC films can also be obtained by chemical vapor deposition (CVD) like WTe$_2$[12] but this technique may result in a high impurity level and structural defects. In contrast, molecular beam epitaxy (MBE) offers a high achievable film quality but remains challenging owing to the different thermodynamic properties of chemical elements used as precursors in TMDCs (adsorption, sticking coefficient, surface diffusion, …). MBE of WTe$_2$ has been reported solely on a reduced number of substrates and only in the few-monolayer (ML) thickness regime[13]. Microscopic information about the nucleation and growth mechanisms of WTe$_2$ is also missing. Moreover, the reported lateral island size of WTe$_2$ grown by MBE remains limited to typically 20-40 nm[14–16]. It is thus necessary to increase the lateral island size of WTe$_2$ thin films to make it suitable for potential device applications. In this work, we report the growth of T$_d$-WTe$_2$ thin films by van der Waals molecular beam epitaxy (MBE) on ML-graphene/SiC substrate from monolayer to >5nm thicknesses. Growth conditions, including temperature and growth rate, have been explored in order to increase the lateral nanoflake size above 100nm. A simple model allowing for the estimation of the island size as a function of the growth parameters, based on nucleation theory[17] and Kolmogorov-Johnson-Mehl-Avrami (KJMA) approach[18–21], is proposed and could be of significance for future studies on van der Waals MBE of TMDCs. Finally, we report the band structure measurements of $WTe_2$ by angular resolved photoemission spectroscopy (ARPES) in agreement with first principle calculations performed on free-standing $WTe_2$.

METHODS

$WTe_2$ thin films have been deposited in a molecular beam epitaxy (MBE) chamber under ultra-high vacuum (UHV) on graphene/SiC(0001) substrate[22] with *in situ* monitoring using RHEED diffraction technique. Post-synthesis characterizations include scanning tunneling microscopy (STM) and angular-resolved photoemission spectroscopy (ARPES). First-principle electronic band structures of ML-$WTe_2$ and bilayer (BL) $WTe_2$ films have been determined using Quantum espresso code[23,24] with Perdew–Burke–Ernzerhof (PBE) exchange correlation functional[25] and full-relativistic ultrasoft pseudopotentials from the PSlibrary.1.0.0[26] (see Section 1 of the Supporting Information for details). Bulk $WTe_2$ exhibits a $T_d$ phase characterized by a layered orthorhombic lattice ($\alpha = 90°$) with inversion symmetry breaking (see Section 2 of the Supporting Information). Growth of $WTe_2$ by MBE is challenging mainly due to the difficulty to incorporate Te. It is due to a sticking coefficient of chalcogenides much lower than that of transition metals[27] and the difficulty to sublimate atomic Te[28]. A lack of Te during the deposition leads to amorphous and non-stoichiometric growth[13]. To circumvent this problem, it is necessary to use a large Te:W flux ratio during the growth. Te:W ratios as high as 800:1 have been employed by Walsh *et al.*[13] along with a W-beam interruption technique. Here, similar flux ratios have been employed along with relatively low deposition temperatures to limit Te desorption and low growth rate of about 1 ML/h to favour atomic diffusion and W-Te bond creation.

EPITAXIAL ELABORATION AND SURFACE MORPHOLOGY

*In-situ* RHEED patterns of substrate and ML-WTe$_2$ film are presented in Figure 1(a). The following epitaxial relationships can be deduced: WTe$_2$ [100] is parallel to graphene [100] and WTe$_2$ [010] is parallel to graphene [$2\bar{1}0$] as reported in Ref. [29]. Lattice constants determined from diffraction patterns reads $a = 3.5 \pm 0.2$Å for the [100] direction and $b = 6.5 \pm 0.3$Å for the [010] direction. These values are consistent with the lattice parameters of bulk T$_d$-WTe$_2$ and previous MBE growth of WTe$_2$[13]. Film with thicknesses >5nm have been successfully investigated. Figure 1(b) shows for instance a transmission electron microscopy (TEM) image along [100] zone axis of a nominally 6nm-WTe$_2$ (8ML). Fast Fourier transform (FFT) of Figure 1(b) confirms T$_d$-WTe$_2$ lattice parameters with $b = 6.35 \pm 0.15$ Å and $c = 14.5 \pm 0.3$ Å. More importantly, a measured angle $\alpha = 90.4 \pm 0.8°$ between the [010] and the [001] directions supports a orthorhombic stacking expected for the T$_d$ phase. The growth of the T$_d$-phase is also confirmed by the vibrational modes identified in the Raman spectrum shown in Figure 2(c)[30,31]. The bands around 150cm$^{-1}$ are attributed to 6H- SiC substrate modes[32].

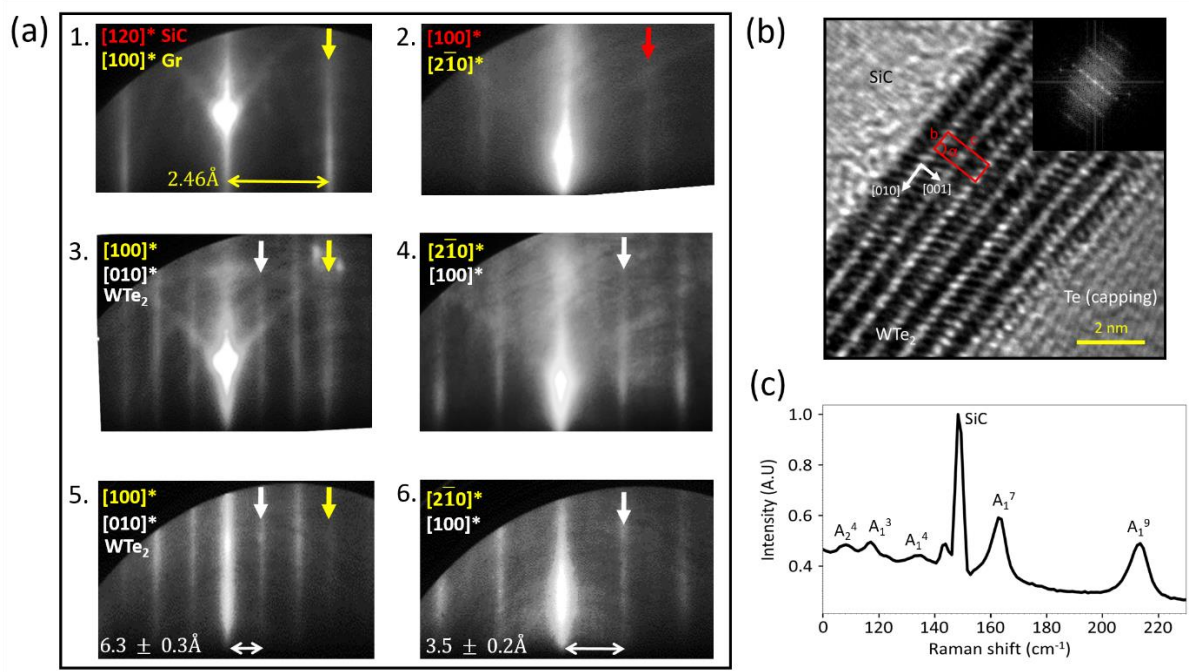

Figure 1. (a) *In-situ* RHEED patterns upon growth of $WTe_2$ on ML-graphene/SiC along $WTe_2$ [010]* (a1, a3, a5) and $WTe_2$ [100]* (a2, a4, a6) reciprocal directions. Images a1 and a2 are graphene/SiC substrate RHEED patterns. Images a3 and a4 (resp. a5 and a6) are $WTe_2$ on graphene/SiC RHEED patterns at the beginning (resp. end) of the growth. The arrows denote the different diffraction streaks. (b) Cross-sectional TEM of a Te-capped 6nm-$WTe_2$ in a [100] zone axis. The inset is the FFT of the TEM image. (c) Raman spectrum of $WTe_2$ film with $T_d$-$WTe_2$ structure vibrational modes (638nm laser excitation).

Surface morphology of a ML-$WTe_2$ film measured by STM is shown in Figure 2(a). For this sample, the mean lateral island size S is 77±17nm as determined from a Gaussian fit of the island size distribution (see bottom inset) and graphene/SiC atomic steps are fully decorated by $WTe_2$ flakes, while uncovered graphene area can still be found on terraces. This suggests that islands nucleate preferentially at step edges. Observations from *in-situ* RHEED diffraction patterns show that when we employ smaller Te:W flux ratios during growth, a polycrystalline growth occurs swiftly and include, according to Refs. [13,33], Te-deficient clusters (see Section 3 of the Supporting Information for details). Only a reduced number of Te-deficient clusters are visible in Figure 2(a) (blue arrow). Height profiles of the first and the second layers are shown in Figure 2(b). The step height between graphene and the first $WTe_2$ layer is $0.85 \pm 0.03$ nm while the step height between the first and the second $WTe_2$ layers is $0.70 \pm 0.02$ nm which corresponds to $T_d$-$WTe_2$ bulk unit cell height[34]. The large step height between the substrate and the first ML suggests a weak van der Waals-like interaction between the two materials in agreement with the results for epitaxial $WTe_2$ on BL-graphene [35]. LEED measurement (Fig. 2(c)) also indicates that the film consists of three equivalents domains rotated by 120° with respect to each other. They correspond to three different orientations of the $WTe_2$ rectangular unit cell on graphene[14]. $WTe_2$ LEED patterns also show an orientation disorder measured of $5.6 \pm 2.4°$.

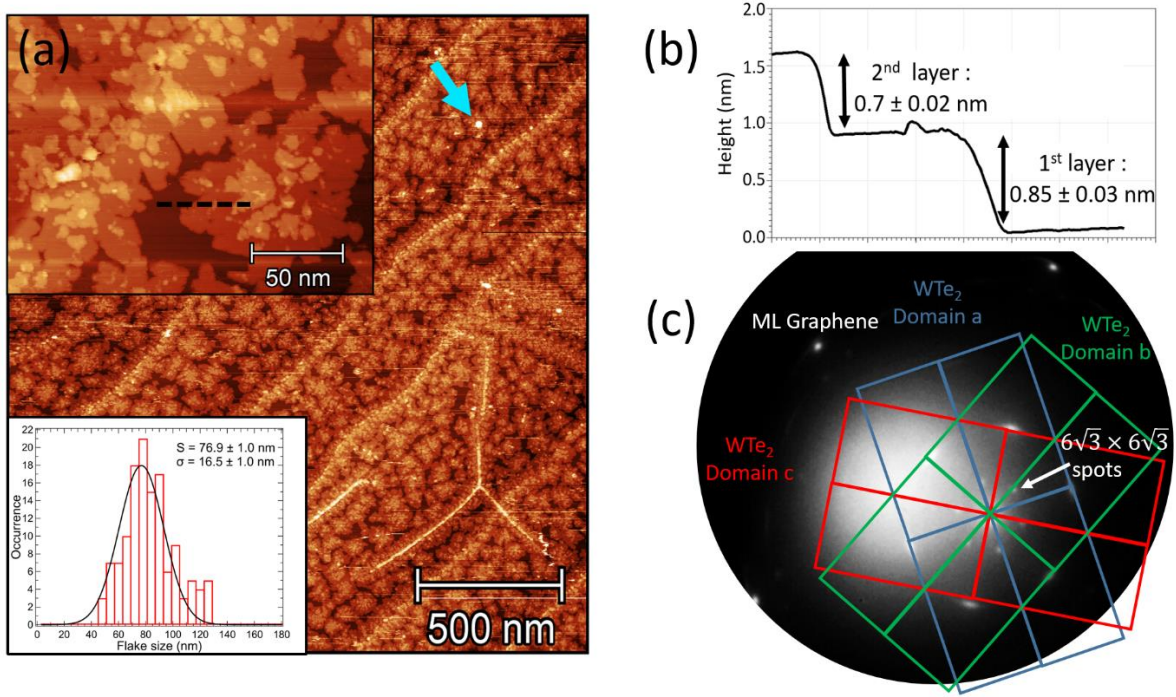

Figure 2. Sub-monolayer WTe$_2$ deposited on graphene/SiC. (a) 2μm × 2μm STM image ($I_t$= 150 pA, U= 1 V) of WTe$_2$ showing monolayer flakes partly covered by bilayers. The top inset shows a 200 nm × 200 nm zoom-in and the bottom inset represents the flake size distribution with a Gaussian fit reading a mean flake size of S=77±17nm. The blue arrow shows a Te-deficient cluster (see text for details). (b) Height profile of WTe$_2$ monolayer and bilayer determined along the dashed line of the inset of (a). (c) ML- WTe$_2$ LEED pattern showing the presence of three domain orientations.

NUCLEATION DENSITY ANALYSIS AND CRITICAL COVERAGE

To have a better grasp on growth conditions of WTe$_2$, the effects of the substrate temperature and deposition fluxes on the nanoflake nucleation density and lateral sizes have been investigated. For all samples, we have used a deposition duration of 1hr. A temperature range between 220°C and 330°C has been covered. Below 220°C, the sticking coefficient of Te is too high, resulting in the deposition of a Te film due to the high Te flux. Above 350°C an unidentified phase appears (see Section 4 of the Supporting Information). The impact of the growth temperature is illustrated in Figure 3(a-c). As summarized in Figure 3(g), increasing the temperature leads to a decrease of the islands nucleation density by promoting the diffusion of adatoms.

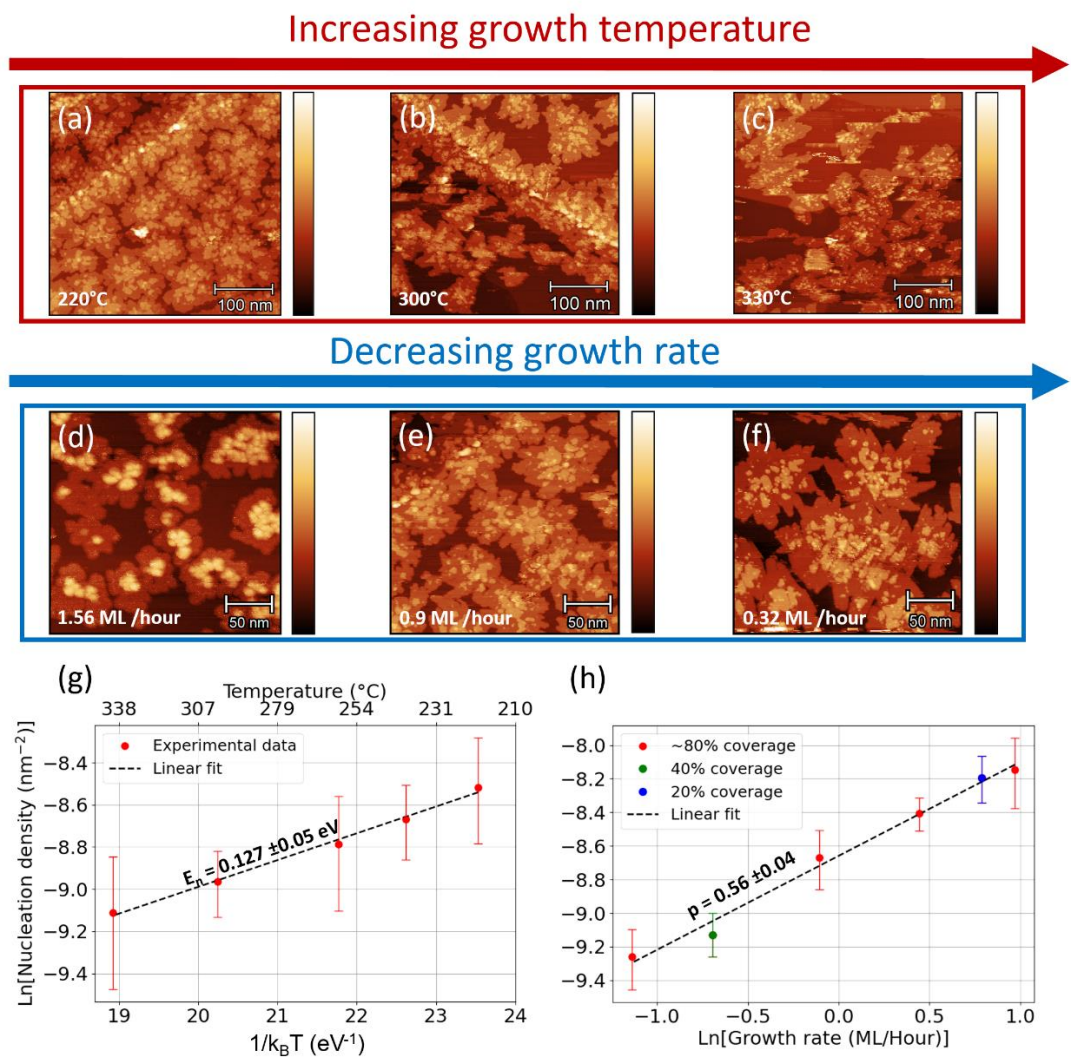

Figure 3. STM images of the WTe$_2$ growth as a function of the temperature (a-c, Te:W ratio 800:1, deposition time: 1h) and the growth rate (d-f, Te:W ratio 800:1, deposition temperature ≃250°C, surface coverage 80%-90%). (g) Arrhenius plot of WTe$_2$ nucleation density. (h) $Ln-ln$ plot of WTe$_2$ nucleation density as a function of the growth rate.

The effect of the growth rate, controlled here by the W flux, has been investigated in the range of 0.32 ML/h to 2.64 ML/h at a growth temperature of ≃250°C for 80%-90% surface coverages (Fig. 3(d–f)). As depicted in Figure 3(h), decreasing the growth rate leads to lower nucleation densities that results in an increase of the island sizes. At 0.32 ML/h, the mean island size is 110 ± 10nm which is more than three time larger than the reported values found in the literature[14–16]. A slower growth rate results from a lower metal flux, which means less adsorbed W atoms on the surface available for flake creation and, in turn, a lower nucleation density.

We now turn to a quantitative analysis of $T_d$-$WTe_2$ growth properties using nucleation theory. This theory was originally developed for single-element depositions and has been recently extended to binary materials with negligible elemental desorption[36]. Te sublimation becomes however significant for surfaces above 200°C[37]. As a consequence, Te adatoms have a very limited lifetime on the surface as compared to W adatoms in our experimental conditions[27]. Secondly, as the Te flux is significantly higher than the W flux, the surface can be viewed as saturated with Te adatoms. The significant Te/W flux imbalance also makes W atoms the limiting available species. For these reasons, we can safely postulate that the species that form, diffuse, and aggregate on the substrate are in fact W atoms surrounded with Te atoms that we will call, in the following analysis, "pseudo-atoms", composed of a single chemical formula (*i.e.* one W atom with two Te atoms). As reviewed in Ref. [17], the surface density of nucleated flake of a given material on a substrate can be expressed as

$$N \approx \eta(\theta) N_0 \left(\frac{R}{\nu N_0}\right)^p e^{\frac{E_n}{k_B T}} \quad (1)$$

where $\eta(\theta)$ is a dimensionless nucleation density at the flake surface coverage fraction $\theta$ ($\eta(\theta) \simeq 0.1$ for $\theta > 0.6$ (*21*)), $N_0$ is the number of adsorption sites per unit area (1.2x10$^7$ μm$^{-2}$, *i.e.* the atomic density of the graphene), $\nu$ is the in-plane atom vibration frequency ($\sim$ 1x10$^{11}$ – 1x10$^{13}$ s$^{-1}$), $R$ is the flux of atoms (here W deposition flux), $p$ is an exponent that depends on the size of the critical cluster $i$ (smallest growing island) ; $k_B$ is the Boltzmann constant ; $T$ is the growth temperature and $E_n$ is an effective activation energy.

From Figure 3(g) and 3(h), we can derive $E_n$ and $p$ of equation (1) with $E_n = 0.127 \pm 0.016 eV$ and $p = 0.56 \pm 0.04$. With these values, we can determine the diffusion energy $E_d$ of WTe$_2$ pseudo-atoms and the critical size $i$ of a stable WTe$_2$ cluster. We here reasonably assume that the critical cluster is 2D and that the condensation is complete (*i.e.* negligible WTe$_2$ pseudo-atom desorption with respect to diffusion and flake growth). In this case, $p = \frac{i}{i+2}$ and $E_n = \frac{E_i + iE_d}{i+2}$, where $E_i$ is the cohesion energy of a flake of size $i$. From $p$ we can derive the smallest stable WTe$_2$ cluster to be of size $i = 2.5 \pm 0.5$. This value is compatible with the size of a single WTe$_2$ surface unit cell that contains two chemical formulas. We can thus infer that the kinetic barrier to overcome for WTe$_2$ nucleation corresponds to the formation of a single surface unit cell. To calculate the diffusion energy $E_d$ from $E_n$, the cohesion energy $E_i$ that depends on the effective binding energy $E_b$ of WTe$_2$ and the flake geometry, is needed. As a rough estimate we can consider a mean value $E_i = 2E_b$ (two bonds between two WTe$_2$ neighboring pseudo-atoms). Using the formation enthalpy of $E_F = 0.39 \pm 0.05$ eV/WTe$_2$[38] with 8 bonds per WTe$_2$ formulas in the bulk, we determine a diffusion energy of WTe$_2$ pseudo-atoms on graphene surface of $E_d \simeq 0.2$ eV. This diffusion energy is rather low for small clusters (see for instance Ref. [39]). It reinforces the previous conclusion regarding WTe$_2$ van der Waals epitaxy on ML-gr. Let us note that Figure 3(h) has been obtained for various coverages with all data points well-aligned. This suggests that nucleation is mainly present during the first stages of the growth and can be neglected above a certain, called critical, coverage as reported for WSe$_2$[33] (here $\theta_c < 0.2$ML).

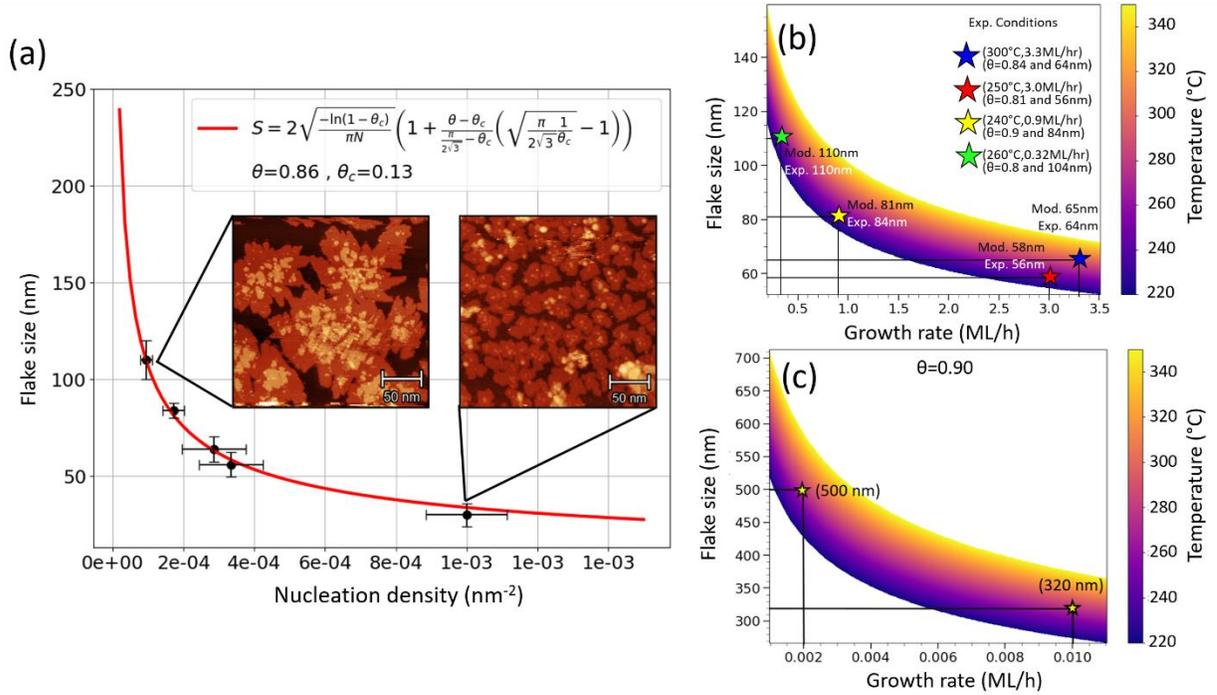

Figure 4. (a) WTe$_2$ flake size as a function of the flake nucleation density. The red curve is the best numerical fit of experimental data obtained with Eq. (2) and $\theta_c = 0.13$ML as unique fitting parameter. (b) Comparison between analytical model and experimental data on the WTe$_2$ flake size. (c) Theoretical WTe$_2$ island size as function of growth rate and growth temperature for a surface coverage of 90% in the extreme slow growth rate regime.

ANALYTICAL MODEL AND FLAKE SIZE EVOLUTION

Beyond the nucleation density analysis, we have also developed an analytical model that predicts the mean lateral size $S$ of WTe$_2$ flakes as a function of the nucleation density $N$ and of the surface coverage $\theta$ using KMJA formalism [18–21]. Let us here mention that these parameters are in fact functions of the deposition conditions (such as temperature and deposition rate). Interestingly enough, our model incorporates the aforementioned critical coverage $\theta_c$ where flake nucleation ends and, by doing so, could be instrumental in understanding microscopic mechanisms at work in TMDC van der Waals epitaxy. More specifically, this approach predicts that the island size reads (see Sec. 5 of Supporting Information for model derivation):

$$S = 2\sqrt{\frac{-\ln(1-\theta_c)}{\pi N}} \left(1 + \frac{\theta - \theta_c}{\frac{\pi}{2\sqrt{3}} - \theta_c} \left(\sqrt{\frac{\pi}{2\sqrt{3}}\frac{1}{\theta_c}} - 1\right)\right) \quad (2)$$

As shown in Figure 4(a), the proposed equation fits well our experimental data with a single fitting parameter, namely $\theta_c$, and an optimized value of 0.13ML. This value agrees perfectly with the values of 0.1-0.15ML reported for WSe$_2$ in Ref. [33]. In addition, this value is in good agreement with our experimental data from Figure 3(h) where an upper limit $\theta_c < 0.2$ML has been inferred. By combining the equation of the nucleation theory[17] and Eq. 2, we can finally estimate the island size as a function of the experimental parameters. Figure 4(b) represents the modelled WTe$_2$ island size and experimental data as a function of the growth rate for different temperatures and a coverage of 86%. The excellent agreement confirms *a posteriori* the validity of our approach. By extrapolating our calculations (Fig. 4(c)), we can expect to further increase the mean flake size by a factor 2-3 by reducing the growth rate while remaining in realistic deposition times (90 hours for 330nm diameter flakes). Larger flake can indeed facilitate detailed studies on individual structures and avoid contributions from multiple domains.

ELECTRONIC PROPERTIES

As a prerequisite for QSHI properties, the opening of a gap in the electronic band structure induced by spin-orbit coupling has been confirmed using angle-resolved photoemission spectroscopy (ARPES) on ML- and BL-WTe$_2$ (see Section 6 of the Supporting Information). The band structure along the $\Gamma - Y$ direction measured on the ML-WTe$_2$ film presented in Figure 2 is shown in Figure 5(a-b). The corresponding DFT calculation performed on free-standing monolayer WTe$_2$ is displayed in Figure 5(c). The band structure along the high symmetry $\Gamma - Y$ path (Fig. 5(d)) is mainly characterized by electron pockets at the Fermi level and a hole-type band slightly below the Fermi level. These features are similar to previous reports found in the literature[14] and validate the crystalline quality of our samples. Let us note that the predicted band gap is underestimated by our DFT calculation with a negative value. This is indeed a known limitation of the PBE functional. With the help of the second derivative of the band structure image (Fig. 5(b)), the band edges, including the band gap, become more discernible. The measured band gap of $56 \pm 11$ meV falls in line with previously reported values[14], and consolidates experimental evidence of the band gap opening due to a strong spin-orbit coupling and band inversion in monolayer WTe$_2$[40,41].

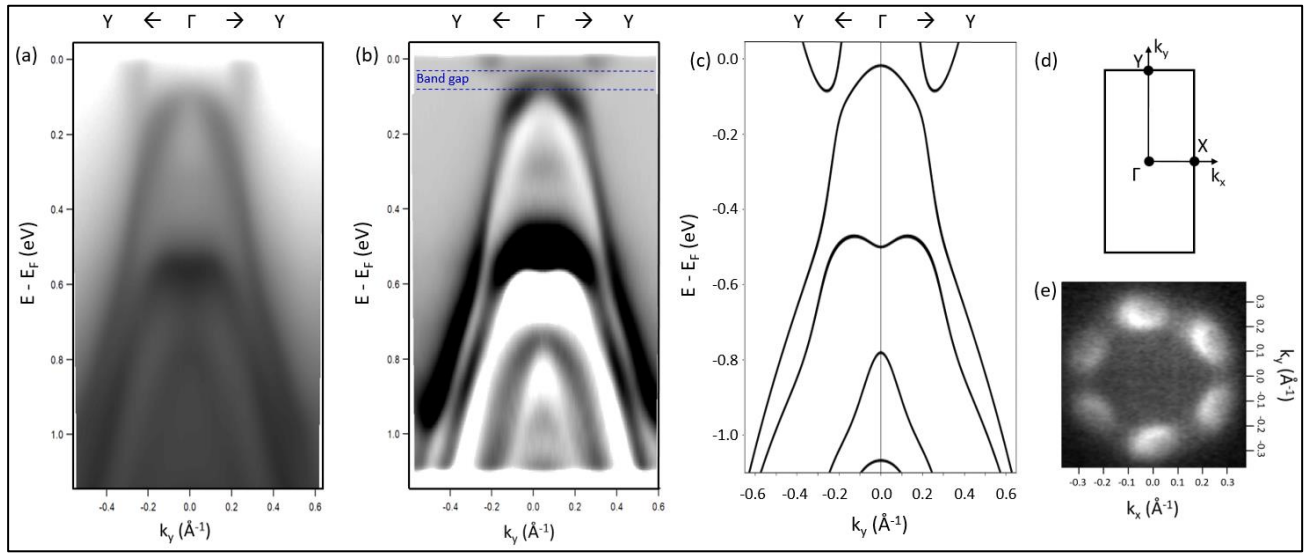

Figure 5. (a) Experimental ML-WTe$_2$ band structure along the $\Gamma$-Y direction. (b) Second derivative of (a). (c) ML-WTe$_2$ bands structure along the $\Gamma$-Y direction calculated by DFT. (d) WTe$_2$ Brillouin zone. (e) Measured experimental Fermi surface determined at 0.03 eV below the Fermi level.

The Fermi surface is characterized by six electron pockets (Fig. 5(e)) owing to the presence of the three orientational domains (two electron pockets per rotational domains). This has been also observed for ML-WTe$_2$ on BL-graphene[14]. The six electron pockets are slightly below the Fermi level in our measurement, with a band minimum located at 46 ± 5meV. We note a significant discrepancy with the literature reports on monolayer WTe$_2$ on BL-graphene[14] regarding the binding energy of the minimum of these electron pockets, which was reported to be less substantial (26 meV below the Fermi level). We attribute this difference to a charge transfer between the WTe$_2$ layer and the underlying graphene substrate resulting in an increased *n*-type doping in our case. This is in agreement with DFT calculations reported in Ref. [29] and has also been observed in WSe$_2$[42]. To estimate the two-dimensional carrier density available for electronic transport, we have assumed a paraboloid dispersion relation, given by $E(k) = \frac{\hbar^2 k_x^2}{2m_x} + \frac{\hbar^2 k_y^2}{2m_y}$ where $m_x$ and $m_y$ are the effective masses respective to the *X* and *Y* directions of the reciprocal space. For ML-WTe$_2$ and BL-WTe$_2$, similar values of $m_x = 0.24 \pm 0.03 m_e$ and $m_y = 0.09 \pm 0.03 m_e$ are derived. In the free-carrier approximation, 2D carrier densities for a single electron pocket are determined to be $n_{ML} = 2.0 \pm 0.5 \times 10^{12}$cm$^{-2}$ and $n_{BL} = 3.1 \pm 0.7 \times 10^{12}$cm$^{-2}$. These values exceed the density measured in Ref. [43] on exfoliated WTe$_2$. This supports the hypothesis of a charge transfer between the WTe$_2$ layer and the graphene substrate which origins remains to be investigated.

CONCLUSION

To summarize, we have elaborated T$_d$-WTe$_2$ on ML-graphene by molecular beam epitaxy from ML to unreported >5nm thicknesses. Sub-ML films show WTe$_2$ flakes with the largest lateral sizes found in the literature (> 100nm). The occurrence of the T$_d$-phase is confirmed by TEM and Raman spectroscopy measurements. We have developed a simple analytical model based on nucleation theory and KJMA equation that estimates WTe$_2$ flake size evolution as a function of experimental parameters. The quality of the samples is attested by the gap opening in the ARPES band structure and the occurrence of electron pockets as expected in the QSHI regime. The experimental band structure is in agreement with DFT calculations of free-standing WTe$_2$. We have estimated the carrier density of WTe$_2$ electrons pocket and have shown that a charge transfer occurs between the monolayer graphene substrate and the WTe$_2$ resulting in a *n*-type doping of the WTe$_2$.

ASSOCIATED CONTENT

**Supporting Information**. Additional details on graphene substrates, elaboration conditions, nucleation model and bilayer WTe$_2$ electronic properties (pdf).

AUTHOR INFORMATION

**Corresponding Author**

*Fabien Cheynis, fabien.cheynis@univ-amu.fr

**Author Contributions**

The manuscript was written through contributions of all authors. All authors have given approval to the final version of the manuscript.

**Funding Sources**

The project leading to this publication has received funding from Excellence Initiative of Aix-Marseille University A*MIDEX and a French "Investissements d'Avenir" program through the AMUtech Institute. This work has also been supported by the ANR grant FETh (Grant No. ANR-22-CE08-0023) and the PACA region project CROISSANT.

ACKNOWLEDGMENT

We thank Martiane Cabié (CP2M, Marseille, France) for lamella preparation of $WTe_2$ thin films by FIB. We also thank Alexandre Altié and Damien Chaudanson (CINaM, Marseille, France) for transmission electron microscopy support.

For Table of Contents Only

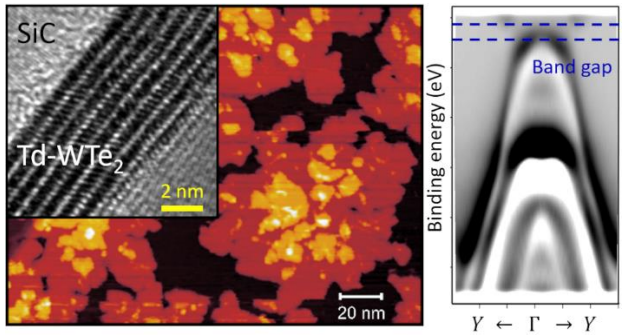